\documentclass[journal]{IEEEtran} 

\usepackage[T1]{fontenc}
\usepackage[latin9]{inputenc}
\usepackage{verbatim}
\usepackage{float}
\usepackage{amstext}
\usepackage{graphicx}
\usepackage{amssymb}
\usepackage{fancybox}
\usepackage{calc}
\usepackage{amsmath}
\usepackage{xfrac}
\usepackage{cite}
\usepackage{flushend}
\makeatletter

\floatstyle{ruled}
\newfloat{algorithm}{tbp}{loa}
\floatname{algorithm}{Algorithm}

\newtheorem{prop}{Proposition}

\newtheorem{alg}{Algorithm}

\renewcommand{\fnum@figure}{Fig.~\thefigure}

\newcommand{\ie}{{\it i.e.}}

\newcommand{\Rset}{{\mathbb{R}}}
\newcommand{\Cset}{{\mathbb{C}}}

\newcommand{\Tr}{\text{Tr}}

\newcommand{\Diag}{{\bf Diag}}

\usepackage{ifpdf} \ifpdf 
\usepackage[unicode=true,
bookmarks=false,bookmarksnumbered=false,bookmarksopen=false,bookmarksopenlevel=2,breaklinks=false,pdfborder={0 0
  0},backref=false,colorlinks=false,citecolor=blue,pdftex] {hyperref}
\else 
\usepackage[unicode=true,
bookmarks=false,bookmarksnumbered=false,bookmarksopen=false,bookmarksopenlevel=2,breaklinks=false,pdfborder={0 0
  0},backref=false,colorlinks=false,citecolor=blue,dvips] {hyperref}
\fi

\hypersetup{pdftitle={Precoding for Relay Networks},
  pdfauthor={W. Zeng},
  pdfsubject={cooperative strategy},
  pdfkeywords={Precoding for Relay Networks}}

\makeatother

\newcommand{\V}{\mathbf{V}}
\newcommand{\MI}{\mathcal{I} \left(\mathbf{x;y}\right)}
\newcommand{\St}{\text{St}(n)}

\newcommand{\vlambda}{{\boldsymbol\lambda}}
\newcommand{\vsigma}{{\boldsymbol\sigma}}

\newcounter{MYtempeqncnt}%

\begin{document}

\title{Linear Precoding for Relay Networks with Finite-Alphabet Constraints}

\author{Weiliang Zeng\IEEEauthorrefmark{1}, Chengshan Xiao\IEEEauthorrefmark{2}, Mingxi Wang\IEEEauthorrefmark{2}, and Jianhua Lu\IEEEauthorrefmark{1}
  \\
  \IEEEauthorrefmark{1}State Key Laboratory on Microwave and Digital
  Communications\\
  Tsinghua National Laboratory for Information Science and Technology\\
  Dept. of Electronic Engineering, Tsinghua University, Beijing,
  China\\
  \IEEEauthorrefmark{2}Dept. of Electrical \& Computer Engineering,
  Missouri University of Science and Technology, MO, USA\\
  Email: \{zengwl, lujh\}@wmc.ee.tsinghua.edu.cn, \{xiaoc, mw9zd\}@mst.edu 
  \thanks{This work was supported in part by the National Basic Research Program of China (2007CB310601) and the U.S. National Science Foundation under Grant CCF-0915846. }
  \thanks{This work has been carried out while W. Zeng is a visiting scholar at Missouri University of Science and Technology.}
}

\maketitle

\begin{abstract}
  In this paper, we investigate the optimal precoding scheme for relay networks with finite-alphabet constraints. We show that the previous work utilizing various design criteria to maximize either the diversity order or the transmission rate with the Gaussian-input assumption may lead to significant loss for a practical system with finite constellation set constraint. A linear precoding scheme is proposed to maximize the mutual information for relay networks. We exploit the structure of the optimal precoding matrix and develop a unified two-step iterative algorithm utilizing the theory of convex optimization and optimization on the complex Stiefel manifold. Numerical examples show that this novel iterative algorithm achieves significant gains compared to its conventional counterpart. 
\end{abstract}

\section{Introduction}

Cooperative relaying is an emerging technology, which provides reliable high data rate transmission for wireless networks without the need of multiple antennas at each node. These benefits can be further exploited by utilizing judicious cooperative design (see \cite{Behbahani2008,gershman2010SPM,khajehnouri2007,Rong2009TSP,Zeng2010Twireless} and the references therein).

Most of the existing designs optimize the performance of relay networks with Gaussian-input assumptions, for example, maximizing output signal-to-noise (SNR) \cite{Behbahani2008,gershman2010SPM}, minimizing mean square error (MSE) \cite{khajehnouri2007,Behbahani2008} and maximizing channel capacity \cite{Behbahani2008,Rong2009TSP,Zeng2010Twireless}. Despite the information theoretic optimality of Gaussian inputs, they can never be realized in practice. Rather, the inputs must be drawn from a finite constellation set, such as pulse amplitude modulation (PAM), phase shift keying (PSK) modulation and quadrature amplitude modulation (QAM), in a practical communication system. These kinds of discrete constellations depart significantly from the Gaussian idealization \cite{lozano2006,Xiao2008Gcom}. Therefore, there exhibits a big performance gap between the scheme designed with the Gaussian-input assumption and the scheme designed from the standing point of finite-alphabet constraint \cite{Zeng2010Gcom}.

In this paper, we consider the two-hop relay networks with finite-input constraint and utilize linear precoder to improve the maximal possible transmission rate of networks. We exploit the optimal structure of the precoding matrix under finite-alphabet constraint and develop a unified framework to solve this nonconvex optimization problem.
We prove that the left singular matrix of the precoder coincides with the right singular matrix of the effective channel matrix; the mutual information is a concave function of the power allocation vector of the precoder; the optimization of the right singular matrix with unitary constraint can be formulated as an unconstrained one on the complex Stiefel manifold. Once these important results are provided, the optimal precoder is solved with a complete two-step iterative algorithm utilizing the theory of convex optimization and optimization on the manifold. We show that this novel iterative algorithm achieves significant gains compared to its conventional counterpart.

\emph{Notation:} Boldface uppercase letters denote matrices, boldface lowercase letters denote column vectors, and italics denote scalars. The superscripts $\left(\cdot\right)^{T}$ and $\left(\cdot\right)^{H}$ stand for transpose and Hermitian operations, respectively.  The subscripts $c_{i}$ and $\left[\mathbf{A}\right]_{ij}$ denote the $i$-th element of vector $\mathbf{c}$ and the ($i, j $)-th element of matrix $\mathbf{A}$, respectively. The operator $ \bf{Diag} (\mathbf{a}) $ represents a diagonal matrix whose nonzero elements are given by the elements of vector $ \mathbf{a} $. Furthermore, $ {\bf vec} \left(\mathbf{A}\right)$ represents the vector obtained by stacking the columns of $ \mathbf{A}$; $\mathbf{I}$ and $ \mathbf{0} $ represents an identity matrix and a zero matrix of appropriate dimensions, respectively; $\Tr\left(\mathbf{A}\right)$ denotes the trace operation. Besides, all logarithms are base 2.

\vspace{-2pt}

\section{System Model}
\label{sec:system-model}

\begin{figure*}[!t]
  \normalsize
  \setcounter{MYtempeqncnt}{\value{equation}}
  \setcounter{equation}{2}
  \begin{equation}
    \label{eq:MutualInfo}
    \MI=
    \log M - \frac{1}{2LM^{2L}}\sum_{m=1}^{M^{2L}}\mathbb{E}_{\mathbf{n}}\left\{
      \log\sum_{k=1}^{M^{2L}}\exp\left[-\Vert\mathbf{HP}\left(\mathbf{x}_{m}-\mathbf{x}_{k}\right)+\mathbf{n}\Vert^{2}+\Vert\mathbf{n}\Vert^{2}\right]\right\}.
  \end{equation}
  \setcounter{equation}{\value{MYtempeqncnt}}
  \hrulefill
  \vspace*{-10pt}
\end{figure*}

\begin{figure*}[!t]
  \normalsize
  \setcounter{MYtempeqncnt}{\value{equation}}
  \setcounter{equation}{8}
  \begin{align}
    \label{eq:MMSE}
    \mathbf{E}
    =\mathbf{I}-\frac{1}{\left(\pi M\right)^{2L}}\int_{\mathbf{y}}\frac{\left[\sum_{l=1}^{M^{2L}}\mathbf{x}_{l}\exp\left(-\Vert\mathbf{y}-\mathbf{HP}\mathbf{x}_{l}\Vert^{2}\right)\right]\left[\sum_{k=1}^{M^{2L}}\mathbf{x}_{k}^{H}\exp\left(-\Vert\mathbf{y}-\mathbf{HP}\mathbf{x}_{k}\Vert^{2}\right)\right]}{\sum_{m=1}^{M^{2L}}\exp\left(-\Vert\mathbf{y}-\mathbf{HP}\mathbf{x}_{m}\Vert^{2}\right)}d\mathbf{y}.
  \end{align}
  \setcounter{equation}{\value{MYtempeqncnt}}
  \hrulefill
  \vspace*{-10pt}
\end{figure*}

Consider a relay network with one transmit-and-receive pair, where the source node attempts to communicate to the destination node with the assistance of $ k $ relays ($r_1, r_2, \cdots, r_k$). All nodes are equipped with a single antenna and operated in half-duplex mode. We consider a flat fading cooperative transmission system, in which the channel gain from the source to the destination is denoted by $ h_0 $, whereas those from the source to the $i$-th relay and from the $i$-th relay to the destination are denoted as $ h_{i} $ and $g_{i}$, respectively.

We focus on the amplify-and-forward protocols combined with single relay selection \cite{Zeng2010Twireless}. The signal transmission is carried out by blocks with block length $ 2L $, $ L\geq1 $. For the selected relay node, there is a data receiving period of length $ L $ before a data transmitting period of length $ L $.

The original data at the source node is denoted by
\begin{equation*}
  \label{eq:trans_data}
  \mathbf{x}=\left[
    \mathbf{x}_{a}^{T} \quad  \mathbf{x}_{b}^{T}
  \right]^T,
\end{equation*}
where $\mathbf{x}_{a}=[x_{1}, \cdots, x_{L}]^{T}$, $\mathbf{x}_{b}=[x_{L+1}, \cdots, x_{2L}]^{T}$ with $ x_{l} $ being the data symbol at the $ l $-th time slot, $ l=1, \cdots, 2L $. We assume that the original information signals are equally probable from discrete signaling constellations such as PSK, PAM, or QAM with unit covariance matrix, \ie, $ \mathbb{E} \left[\mathbf{x}\mathbf{x}^{H}\right]= \mathbf{I}$. 

The original data is processed by a precoding matrix before being transmitted from the source node. The precoded data $\mathbf{s}=\left[ \mathbf{s}_{a}^{T}\;\; \mathbf{s}_{b}^{T}\right]^T$ is given by $\mathbf{s}= \mathbf{Px}$, where $ \mathbf{P}$ is a \emph{generalized} complex precoding matrix.

The source node sends the signal $\sqrt{P_{s}}\mathbf{s}_{a}$ with power constraint $ P_{s} $ during the first $ L $-time slots. Let $\mathbf{y}_{i}$ and $\mathbf{y}_{a}$ be received signals at the $i$-th relay node and the destination, respectively, which are given by
\[
  \mathbf{y}_{i} =\sqrt{P_{s}}h_{i}\mathbf{s}_{a}+\mathbf{n}_{i}, \quad\quad
  \mathbf{y}_{a} =\sqrt{P_{s}}h_{0}\mathbf{s}_{a}+\mathbf{n}_{a},
\]
where $\mathbf{n}_{i}$ and $\mathbf{n}_{a}$ denote, respectively, the independent and identically distributed (i.i.d.) zero-mean circularly Gaussian noise with unit variance at the $i$-th relay and the destination.

Assuming the $ i $-th relay node is selected at the second $ L$-time slots, it scales the received
signal by a factor $ b $ (so that its average transmit power is $ P_r $) and forwards it to the
receiver. We assume only the second-order statistics of $ h_i $ is known at the $ i $-th relay node,
then $ b $ can be chosen as $b=\sqrt{LP_r/\Tr\left(\mathbb{E}\left[\mathbf{y}_{i}\mathbf{y}_{i}^{H}\right]\right)}$.  At the second $ L$-time slots, the source node sends the signal $\sqrt{P_{s}}\mathbf{s}_{b}$. Therefore, the
destination node receives a superposition of the relay transmission and the source transmission
 according to
\begin{align*}
  \mathbf{y}_{b}&=bg_{i}\mathbf{y}_{i}+\sqrt{P_{s}}h_{0}\mathbf{s}_{b}+\mathbf{n}_{b}\nonumber \\
  & =\sqrt{P_{s}}bh_{i}g_{i}\mathbf{s}_{a}+\sqrt{P_{s}}h_{0}\mathbf{s}_{b}+\mathbf{n}_{e},
\end{align*}
where $ \mathbf{n}_{b}$ denotes the noise vector of the destination at the second $ L$-time slots, and $ \mathbf{n}_{e}$ denotes the effective noise $\sim\mathcal{CN}\left(\mathbf{0},N_{d}\mathbf{I}\right)$ with $N_{d}=1+b^2|g_i|^2$.

For convenience in the presentation, we normalize $\mathbf{y}_{b}$ by $w=1/\sqrt{N_{d}}$ and denote the received signal vector as $\mathbf{y}=\left[\mathbf{y}_{a}^{T}\;\; w \mathbf{y}_{b}^{T}\right]^{T}$. Thus, the effective input-output relation for the two-hop transmission with precoding is summarized as
\begin{equation}
  \label{eq:ChannelModel}
  \mathbf{y} = \mathbf{Hs+n} = \mathbf{HPx+n},
\end{equation}
where $\mathbf{x}$ is the original transmitted signal vector;
$\mathbf{n}=\left[\mathbf{n}_{a}^{T}\;\; w \mathbf{n}_{e}^{T}\right]^{T}$ is i.i.d. complex Gaussian channel noise vector with zero mean and unit variance, \ie,
$\mathbf{n}\sim\mathcal{CN}\left(\mathbf{0},\mathbf{I}\right)$; $\mathbf{H}$ is the effective channel matrix of the two-hop relay channel
\begin{equation}
  \label{eq:channelH}
  \mathbf{H}=\sqrt{P_s}
  \left[
    \begin{array}{cc}
      h_{0}\mathbf{I} & \mathbf{0}  \\
      wbh_{i}g_{i} \mathbf{I} & wh_{0} \mathbf{I}
    \end{array}
  \right].
\end{equation}

Our precoding scheme is thus the design of matrix $ \mathbf{P}$ to maximize the mutual information with finite-alphabet constraints. Note that for the proposed algorithm to be effectively implemented in practice, the destination estimates effective channel matrices of relay networks through pilot assisted channel estimation. Then the destination node selects one relay for cooperation and provides the corresponding effective channel to the source node via a feedback channel. Considering the special structure of the channel matrix \eqref{eq:channelH}, the amount of the feedback can be very small. After signal feedback, the source node utilizes the proposed precoding algorithm to optimize the network performance. 

\section{Mutual Information for Relay Networks}
\label{sec:problem-statement}

We consider the conventional equiprobable discrete constellations such as $M$-ary PSK,
PAM, or QAM, where $M$ is the number of points in the signal constellation. The mutual information
between $\mathbf{x}$ and $\mathbf{y}$, with the equivalent channel matrix  $\mathbf{H}$ and the precoding matrix $\mathbf{P}$ known at the receiver, is $\MI$ given by (\ref{eq:MutualInfo}), where $ \Vert \cdot \Vert$ denotes Euclidean norm of a vector; both $\mathbf{x}_{m}$ and $ \mathbf{x}_{k}$ contain $ 2L $ symbols, taken independently from the $M$-ary signal constellation \cite{Xiao2008Gcom,Zeng2010Gcom}. 

\setcounter{equation}{3}

\begin{prop}
  \label{prop:mutual_info}
  Let $ \mathbf{U} $ be a unitary matrix, and the following relationships hold:
  \begin{align}
    \label{eq:mutual_a}
    \mathcal{I} \left(\mathbf{x};\mathbf{y}\right)
    & = \mathcal{I}\left(\mathbf{x; \mathbf{Uy}}\right) \\
    \label{eq:mutual_b}
    \mathcal{I} \left(\mathbf{x};\mathbf{y}\right)
    & \neq \mathcal{I}\left(\mathbf{Ux; \mathbf{y}}\right).
  \end{align}
\end{prop}

\begin{IEEEproof}
  See proof in \cite{Zeng2011TWC}.
\end{IEEEproof}

Proposition \ref{prop:mutual_info} implies that the property of mutual information for the discrete
input vector is different from the case of Gaussian inputs. For Gaussian inputs, the mutual
information is unchanged when either
transmitted signal $ \mathbf{x} $ or received signal $ \mathbf{y} $ is rotated by a unitary matrix. The case of finite inputs does not follow the same rule. Therefore, it provides us a new opportunity to improve the system performance.

The optimization of the linear precoding matrix $ \mathbf{P} $ is carried out over all $ 2L\times2L $ complex precoding matrices under transmit power constraint, which can be cast as a constrained nonlinear optimization problem
\begin{equation}
  \label{eq:Opt-prob-original}
  \begin{array}{ll}
    \mbox{maximize} & \MI\\
    \mbox{subject to} & \Tr\left\{\mathbb{E}\left[\mathbf{ss}^H\right]\right\}
    =\Tr\left(\mathbf{PP}^H\right) \leq 2L.
\end{array}  
\end{equation}

\begin{prop}[Necessary Condition]
  \label{prop:OptimalCondition}
  The optimal precoding matrix $ \mathbf{P}^\star $ for the optimization problem \eqref{eq:Opt-prob-original} satisfies the following condition \cite{Xiao2008Gcom,Zeng2010Gcom}
  \begin{equation}
    \label{eq:nec_cond}
    \mu \mathbf{P}^\star = \mathbf{H}^H \mathbf{HP}^\star \mathbf{E},
  \end{equation}
  where $ \mu  $ is chosen to satisfy the power constraint, and $\mathbf{E}$ is the minimum mean square error (MMSE) matrix given by (\ref{eq:MMSE}).
\end{prop}
\setcounter{equation}{9}

\begin{IEEEproof}
  See proof in \cite{Zeng2011TWC}.
\end{IEEEproof}

It is important to note that Proposition \ref{prop:OptimalCondition} gives a necessary condition satisfied by any critical points, since the optimization problem (\ref{eq:Opt-prob-original}) is nonconvex. It is possible to develop an  algorithm based on the gradient of the Lagrangian as proposed in \cite{palomar2006a,Zeng2010Gcom}. However, this kind of algorithms can be stuck at a local maximum, which is influenced by the starting point and may be far away from the optimal solution. This fact will be shown via an example in the sequel.

\section{Precoder Design to Maximize the Mutual Information}
\label{sec:precoder-design}

\subsection{Optimal Left Singular Vector}
\label{sec:left-singular-matrix}

We start by characterizing the dependence of mutual information $ \MI$ on precoder $ \mathbf{P}$. Consider the singular value decomposition (SVD) of the $ 2L\times 2L $ channel matrix $ \mathbf{H}=\mathbf{U}_{\mathbf{H}}\Diag(\boldsymbol\sigma)\mathbf{V}_{\mathbf{H}}^{H} $, where $ \mathbf{U}_{\mathbf{H}} $ and $ \mathbf{V}_{\mathbf{H}} $ are unitary matrices, and the vector $ \vsigma $ contains nonnegative entries in decreasing order. Note that the equivalent channel matrix $ \mathbf{H} $ defined in \eqref{eq:channelH} is full rank for any nonzero channel gain $ h_{0} $. We also consider the SVD of the precoding matrix $\mathbf{P}=\mathbf{U}_{\mathbf{P}}\Diag(\sqrt{\vlambda})\mathbf{V}_{\mathbf{P}}^{H}$ and define $\mathbf{U}=\mathbf{U}_{\mathbf{P}}$ and $\mathbf{V}=\mathbf{V}_{\mathbf{P}}^{H}$, where $ \mathbf{U} $ and $ \mathbf{V} $ are named as the left and right singular vectors, respectively; the vector $ \vlambda $ is nonnegative constrained by transmit power. 

\begin{prop}
  \label{prop:LeftSingularMatrix}
  The mutual information depends on the precoding matrix $ \mathbf{P} $ only through $ \mathbf{P}^{H}\mathbf{H}^{H}\mathbf{HP} $. For a given $ \mathbf{P}^{H}\mathbf{H}^{H}\mathbf{HP} $, we can always choose the precoding matrix of the form $ \mathbf{P}=\mathbf{V}_{\mathbf{H}}\Diag(\sqrt{\vlambda})\mathbf{V} $ in order to minimize the transmit power  $ \Tr(\mathbf{PP}^{H}) $, \ie, the left singular vector of $ \mathbf{P} $ coincides with the right singular vector of $ \mathbf{H} $. 
\end{prop}

\begin{IEEEproof}
  See proof in \cite{Zeng2011TWC}.
\end{IEEEproof}

From the results in Proposition \ref{prop:mutual_info} and \ref{prop:LeftSingularMatrix}, it is possible to simplify the channel matrix \eqref{eq:ChannelModel} to
\begin{equation}
  \label{eq:simplified-model}
  \mathbf{y} =  \Diag(\vsigma)\Diag(\sqrt{\vlambda})\mathbf{V}\mathbf{x} + \mathbf{n}.
\end{equation}
Now our discussion will be based on this simplified channel model \eqref{eq:simplified-model}. The optimization variables are power allocation vector $\vlambda$ and right singular vector $ \mathbf{V}$, which are the focuses of the next two subsections. In the sequel, we will use $ \mathcal{I}(\vlambda)$ and $\mathcal{I}(\mathbf{V})$ to emphasize the dependence of mutual information on variables $ \vlambda$ and $ \mathbf{V}$, respectively.

\subsection{Optimal Power Allocation}
\label{sec:optim-power-alloc}

Given a right singular vector of the precoder, we consider the following optimization problem over the power allocation vector $\vlambda$  
\begin{equation}
  \label{eq:Opt-prob-lambda}
  \begin{array}{ll}
    \mbox{maximize} & \mathcal{I}(\vlambda) \\
    \mbox{subject to} & \mathbf{1}^{T}\vlambda\leq 2L \\
    & \vlambda \succeq \mathbf{0},
  \end{array}  
\end{equation}
where $ \mathbf{1}$ denotes a column vector with all entries one.

\begin{prop}
  \label{prop:Hessian-concavity}
  The mutual information is a concave function of the squared singular values of the precoder, $ \vlambda$, \ie, the mutual information Hessian with respect to the power allocation vector satisfies $\mathcal{H}_{\vlambda}\mathcal{I}\left(\vlambda\right)\preceq \mathbf{0}$.  Moreover, the Jacobian of the mutual information with respect to the power allocation vector $ \vlambda$ is given by
  \begin{equation}
    \label{eq:propjacobian}
    \nabla_{\vlambda}\mathcal{I}(\vlambda)
    = \mathbf{R}\cdot{\bf vec} \left(\Diag^{2}(\vsigma)\mathbf{VEV}^{H}\right),
  \end{equation}
  where $ \mathbf{R}\in \mathbb{R}^{2L\times 4L^{2}}$ is a reduction matrix given by
  \begin{equation}
    \label{eq:reduction-matrix}
    \left[\mathbf{R}\right]_{i,2L(j-1)+k}=\delta_{ijk}, \quad {i,j,k}\in \left[1,2L\right].
  \end{equation}
\end{prop}

\begin{IEEEproof}
  See proof in \cite{Zeng2011TWC}.
\end{IEEEproof}

The concavity result in Proposition \ref{prop:Hessian-concavity} extends the Hessian and concavity results in \cite[Theorem 5]{Payaro2009TIT} from real-valued signal model to a generalized complex-valued case. It ensures to find the global optimal power allocation vector given a right singular vector $ \mathbf{V}$, and the gradient result in \eqref{eq:propjacobian} provides the possibility to develop a steepest descent type algorithm to achieve the global optimum \cite{boyd2004}. 

We first rewrite the problem \eqref{eq:Opt-prob-lambda} using the \emph{barrier} method:
\begin{equation}
  \label{eq:Opt-prob-lambda-2}
  \begin{array}{ll}
    \mbox{minimize} & f(\vlambda)=-\mathcal{I} \left(\vlambda\right)+\sum\limits_{i=1}^{2L}\phi(-\lambda_{i})+\phi(\mathbf{1}^{T}\vlambda- 2L),
\end{array}  
\end{equation}
where $ \phi(u)$ is the \emph{logarithmic barrier} function, which approximates an indicator illustrating whether constraints are violated
\begin{equation}
  \label{eq:log-barrier}
  \phi(u) =
\begin{cases}
  -(1/t)\log(-u),& u<0 \\
  +\infty,& u\geq0
\end{cases}
\end{equation}
with the parameter $t>0$ setting the accuracy of the approximation. The gradient of objective function \eqref{eq:Opt-prob-lambda-2} is
\begin{equation}
  \label{eq:gradient-f}
  \nabla_{\vlambda} f(\vlambda)=-\mathbf{R}\cdot{\bf vec} \left(\Diag^{2}(\vsigma)\mathbf{VEV}^{H}\right)
  -\frac{1}{t} \left(\mathbf{q}-\frac{1}{2L-\mathbf{1}^{T}\vlambda}\right),
\end{equation}
where $ q_{i}=1/\lambda_{i}$ is the $ i$-th element of vector $ \mathbf{q}$. Therefore, the steepest descent direction is chosen as
\[
\Delta\vlambda= - \nabla_{\vlambda} f(\vlambda).
\]
Then it is necessary to decide a positive step size $ \gamma\in \Rset$ so that $ f(\vlambda+\gamma\Delta\vlambda)<f(\vlambda)$. The backtracking line search conditions \cite{boyd2004} states that $ \gamma$ should be chosen to satisfy the inequalities
\begin{align}
  \label{eq:line-search-1}
  f(\vlambda) -  f(\vlambda+\gamma\Delta\vlambda) & \geq \frac{1}{2}\gamma \Vert \Delta\vlambda \Vert ^{2},\\
  \label{eq:line-search-2}
  f(\vlambda) -  f(\vlambda+2\gamma\Delta\vlambda) & < \gamma \Vert \Delta\vlambda \Vert ^{2}.
\end{align}
The above ideas can be summarized as the following algorithm, which ensures to converge to the optimal power allocation vector because of the concavity.

\begin{alg}
  \label{alg:power-allocation}
  Steepest Descent to Maximize the Mutual Information Over Power Allocation Vector
  \begin{enumerate}
  \item Given a feasible $ \vlambda$, $ t:=t^{(0)}>0$, $ \alpha>1$, tolerance $ \epsilon>0$.
  \item \label{item:gradient}Compute the gradient of $ f$ at $ \vlambda$, $ \nabla_{\vlambda} f(\vlambda)$, as \eqref{eq:gradient-f} and the descent direction $ \Delta\vlambda= - \nabla_{\vlambda} f(\vlambda)$. Set the step size $ \gamma:=1$.
  \item Evaluate $ \Vert \Delta\vlambda\Vert^{2}$. If it is sufficiently small, then go to Step \ref{item:update-t}.
  \item \label{item:line-search-2} If $ f(\vlambda) -  f(\vlambda+2\gamma\Delta\vlambda) \geq \gamma \Vert\Delta\vlambda\Vert^{2}$, then set $ \gamma:=2\gamma$, and repeat Step \ref{item:line-search-2}.
  \item \label{item:line-search-1} If $ f(\vlambda) -  f(\vlambda+\gamma\Delta\vlambda)  < \frac{1}{2}\gamma \Vert\Delta\vlambda\Vert^{2}$, then set  $ \gamma:=\tfrac{1}{2}\gamma$, and repeat Step \ref{item:line-search-1}.
  \item Set $ \vlambda:=\vlambda+\gamma\Delta\vlambda$. Go to Step \ref{item:gradient}.
  \item \label{item:update-t} Stop if $ 1/t<\epsilon$, else $ t:=\alpha t$, and go to step \ref{item:gradient}.
  \end{enumerate}
\end{alg}

\subsection{Optimization Over Right Singular Vector}
\label{sec:optim-over-right}

This section considers an alternative optimization problem for maximizing the mutual information over the right singular vector $ \mathbf{V}$ for a given power allocation vector,
\begin{equation}
  \label{eq:Opt-prob-V}
  \begin{array}{ll}
    \mbox{maximize} & \mathcal{I}(\mathbf{V})\\
  \mbox{subject to} & \V^{H}\V=\V\V^{H}=\mathbf{I}.
\end{array}  
\end{equation}
This unitary matrix constrained problem can be formulated as an unconstrained one in a constrained search space 
\begin{equation}
  \label{eq:Opt-prob-V-2}
  \begin{array}{ll}
    \mbox{minimize} & g(\V)
  \end{array}  
\end{equation}
where we define the function $ g(\V)$ as $ -\mathcal{I}(\V)$, and with domain restricted to the feasible set:
\begin{equation}
  \label{eq:Opt-prob-V-dom}
  \text{dom}\; g= \left\{\V\in\St\right\},
\end{equation}
in which the set $\St$ is \emph{complex Stiefel manifold} \cite{Manton2002TSP}
\begin{equation}
  \label{eq:st-set}
  \St= \left\{\mathbf{V}\in \Cset^{n\times n}|\mathbf{V}^{H}\mathbf{V}=\mathbf{I}\right\}.
\end{equation}

Associated with each point $ \mathbf{V}\in \text{St}(n)$ is a vector space called \emph{tangent space}, which is formed by all the tangent vectors at the point $ \mathbf{V}$.

\begin{prop}
  \label{prop:gradient-on-tangent}
  The gradient of the mutual information on the tangent space is
  \begin{align}
    \label{eq:gradient-on-tangent}
    \nabla_{\V} g(\V)=-\Diag^{2}&(\vsigma) \Diag(\vlambda)\V \mathbf{E} \nonumber\\ 
    &+\V \mathbf{E}\V^{H}\Diag^{2}(\vsigma) \Diag(\vlambda)\V.
  \end{align}
\end{prop}

\begin{IEEEproof}
  See proof in \cite{Zeng2011TWC}.
\end{IEEEproof}

Utilizing the gradient on the tangent space as the descent direction has been suggested in \cite{Manton2002TSP}, \ie, $\Delta\V= - \nabla_{\V} g(\V)$; however, moving towards the direction on the tangent space may lost the unitary property. Therefore, it needs to be restored in each step via projection.

The projection of an arbitrary matrix $ \mathbf{W}\in\Cset^{n\times n}$ onto the Stiefel manifold $ \pi(\mathbf{W})$ is defined to be the point on the Stiefel manifold closest to $ \mathbf{W}$ in the Euclidean norm
\begin{equation}
  \label{eq:Projection}
  \pi(\mathbf{W})=\arg \min_{\mathbf{Q}\in\text{St}(n)}\Vert \mathbf{W-Q}\Vert ^{2}.
\end{equation}

\begin{prop}[Projection]
  \label{prop:projection}
  Let $ \mathbf{W}\in\Cset^{n\times n}$ be a full rank matrix. If the SVD of $ \mathbf{W}$ is $ \mathbf{W}=\mathbf{U_{W}\Sigma V_{W}}$, the projection is unique, which is given by $ \pi(\mathbf{W})=\mathbf{U_{W}}\mathbf{V}_{\mathbf{W}}$.
\end{prop}

\begin{IEEEproof}
  See proof in \cite{Zeng2011TWC}.
\end{IEEEproof}

Combining the search direction and projection method with the line search conditions in \eqref{eq:line-search-1} and \eqref{eq:line-search-2}, we are able to develop the optimization algorithm to maximize the mutual information over the right singular vector $ \V$.

\begin{alg}
  \label{alg:right-singular-matrix}
  Steepest Descent to Maximize the Mutual Information on Complex Stiefel Manifold
  \begin{enumerate}
  \item Given a feasible $ \V\in\Cset^{n\times n}$ such that $ \V^{H}\V=\mathbf{I}$.
  \item \label{item:gradient-V}Compute the gradient of $ g$ at $ \V$, $ \nabla_{\V} g(\V)$, as \eqref{eq:gradient-on-tangent} and the descent direction $ \Delta\V= - \nabla_{\V} g(\V)$. Set the step size $ \gamma:=1$.
  \item Evaluate $\Vert \Delta\V\Vert^{2}=\Tr\{(\Delta\V)^{H}\Delta\V\}$. If it is sufficiently small, then stop.
  \item \label{item:line-search-V-2} If $ g(\V) -  g(\pi(\V+2\gamma\Delta\V)) \geq \gamma \Vert \Delta\V \Vert^{2}$, then set $ \gamma:=2\gamma$, and repeat Step \ref{item:line-search-2}.
  \item \label{item:line-search-V-1} If $ g(\V) -  g(\pi(\V+\gamma\Delta\V))  < \frac{1}{2}\gamma \Vert \Delta\V\Vert^{2}$, then set  $ \gamma:=\tfrac{1}{2}\gamma$, and repeat Step \ref{item:line-search-V-1}.
  \item Set $ \V:=\V+\gamma\Delta\V$. Go to Step \ref{item:gradient-V}.
  \end{enumerate}
\end{alg}

\subsection{Two-Step Approach to Optimize Precoder}
\label{sec:two-step-approach}
Now we are ready to develop a complete two-step approach to maximize the mutual information over a generalized precoding matrix $ \mathbf{P}$ via combining Proposition  \ref{prop:LeftSingularMatrix} and Algorithm \ref{alg:power-allocation} and \ref{alg:right-singular-matrix}. 

\begin{alg}
  \label{alg:general-P}
  Two-Step Algorithm to Maximize the Mutual Information Over a Generalized Precoding Matrix
  \begin{enumerate}
  \item Set the left singular vector of the precoder $ \mathbf{U}:=\mathbf{V_{H}}$, and give a feasible $ \vlambda$ and $ \V$.
  \item \label{item:algFull-alg1} \emph{Update power allocation vector}: Run Algorithm \ref{alg:power-allocation} given $ \V$.
  \item \emph{Update right singular vector}: Run Algorithm \ref{alg:right-singular-matrix} given $ \vlambda$.
  \item Go to Step \ref{item:algFull-alg1} until convergence.
  \end{enumerate}
\end{alg}

\section{Applications}
\label{sec:applications}

We consider a single-relay network with the block length $ L = 1$ and the channel coefficient $h_{0}=0.4$, $h_{1}=1.2$ and $g_{1}=-0.9 \jmath$. We assume the same transmit power at the source and relay node (\ie, $P_{s}=P_{r}=P$), and the SNR is 3 dB. when the elements of the transmitted signal $ \mathbf{x}$ is drawn independently from BPSK constellations, the mutual information is bounded by 1 bit/s/Hz as shown in \eqref{eq:MutualInfo}.

\begin{figure}
  \begin{centering}
    \includegraphics[clip,width=3.2in]{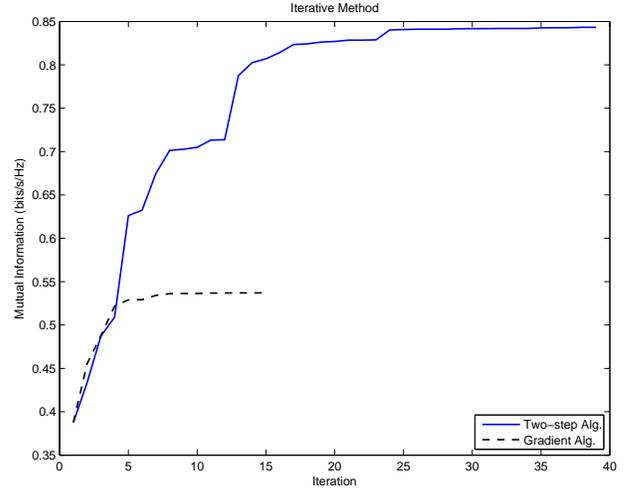} \par
  \end{centering}
  \caption{ Evolution of the mutual information as the linear precoder is iteratively optimized with the two-step algorithm and gradient algorithm.}
  \label{fig:TwoStepConvergence}
\end{figure}

The convergence of the proposed approach is illustrated in Fig. \ref{fig:TwoStepConvergence}. We also show the convergence of algorithms proposed in \cite{Zeng2010Gcom,palomar2006a}. From Fig. \ref{fig:TwoStepConvergence}, it is shown that the direct gradient method is stuck at a local maximum (0.53 bit/s/Hz). The reason for such performance is that the optimization problem is not convex in general. In contrast, the proposed two-step algorithm exploits the characterization of the optimal solution, which leads to a solution with the optimal left singular vector, the optimal power allocation vector (for a given right singular vector), and the local optimal right singular vector from an arbitrary start point. Hence, the algorithm is able to converge to a much higher value (0.85 bit/s/Hz) with about 60 percent improvement. Note that the progress of the proposed method has a staircase shape, with each stair associated with either the iteration for $ t$, named as outer iteration in \cite{boyd2004}, or the change between the optimizations of the power allocation vector and the right singular vector.

\begin{figure}
  \begin{centering}
    \includegraphics[clip,width=3.2in]{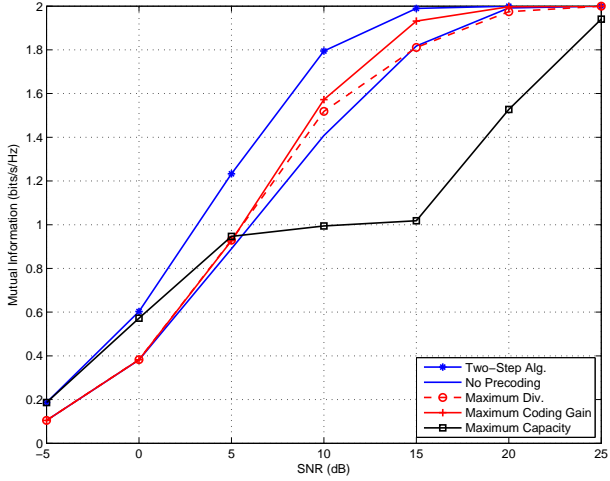} \par
  \end{centering}
  \caption{Mutual Information of relay networks with QPSK inputs.}
  \label{fig:TwoStepSNR}
\end{figure}

The performance of the proposed algorithm is shown is Fig. \ref{fig:TwoStepSNR}, in which the information symbol $\mathbf{x}$ is modulated as QPSK, and the channel is the same as the above case. For the sake of completeness, we also show the performance corresponding to the case of no precoding, maximum diversity design in \cite{ding2007TSP}, maximum coding gain design in \cite{ding2007TSP,ding2008TWC}, and maximum capacity design with Gaussian inputs assumption in \cite{Rong2009TSP}. From Fig. \ref{fig:TwoStepSNR}, we have following several observations.

The method based on maximizing capacity with Gaussian-input assumption may result in a significant \emph{loss} for discrete inputs, especially when the SNR is in medium-to-high regions. The reason comes from the difference in design power allocation vector and right singular vector. For Gaussian inputs, it is always helpful to allocation more power to the stronger subchannels and less power to the weaker subchannels to maximize the capacity. However, this does not work for the case of finite inputs. Since the mutual information of the relay network is upper bounded by $ \log M$ from \eqref{eq:MutualInfo}, there is little incentive to allocate more power to subchannels when they are already close to saturation. Moreover, the right singular vector for Gaussian inputs is an arbitrary unitary matrix to maximize the capacity, because the mutual information is unchanged when the input signal is rotated by a unitary matrix.

The maximum coding gain design has better performance than the method of maximum diversity order and no precoding. We should note that the maximum coding design in \cite{ding2007TSP} is only valid for the case of block length $ L = 1$ and QPSK inputs; it is extended to the case of $ L = 1$ and 16-QAM inputs in \cite{ding2008TWC}.

The proposed two-step precoder optimization results in significant gain on mutual information in a wide range of SNR region. For example, it is about 2 dB, 4 dB and 10 dB better than the method of maximum coding gain, no precoding and maximum capacity, respectively, when the channel coding rate is 2/3. Moreover, this algorithm is able to be utilized for an arbitrary block length $ L$ and input type.

\section{Summary and Conclusion}
\label{sec:summary-conclusion}

In this paper, we have studied the precoding design for dual-hop AF relay networks. In contrast with the previous work utilizing various design criteria with the idealistic Gaussian-input assumptions, we have formulated the linear precoding design from the standpoint of discrete-constellation inputs. To develop an efficient precoding design algorithm, we have chosen the mutual information as the utility function. Unfortunately, the maximization of this utility function over all possible complex precoding matrix is nonconvex, \ie, the direct optimization on the precoder can be stuck at a local maxima, which is influenced by the starting point and may be far away from the optimal solution. We have exploited the structure of the precoding matrix under finite-alphabet constraint and developed a unified framework to solve this nonconvex optimization problem. We have proposed a two-step iterative algorithm to maximize the mutual information. Numerical examples have shown substantial gains of our proposed approach on mutual information compared to its conventional counterparts.


\begin{thebibliography}{10}
  \providecommand{\url}[1]{#1}
  \csname url@samestyle\endcsname
  \providecommand{\newblock}{\relax}
  \providecommand{\bibinfo}[2]{#2}
  \providecommand{\BIBentrySTDinterwordspacing}{\spaceskip=0pt\relax}
  \providecommand{\BIBentryALTinterwordstretchfactor}{4}
  \providecommand{\BIBentryALTinterwordspacing}{\spaceskip=\fontdimen2\font plus
    \BIBentryALTinterwordstretchfactor\fontdimen3\font minus
    \fontdimen4\font\relax}
  \providecommand{\BIBforeignlanguage}[2]{{%
      \expandafter\ifx\csname l@#1\endcsname\relax
      \typeout{** WARNING: IEEEtran.bst: No hyphenation pattern has been}%
      \typeout{** loaded for the language `#1'. Using the pattern for}%
      \typeout{** the default language instead.}%
      \else
      \language=\csname l@#1\endcsname
      \fi
      #2}}
  \providecommand{\BIBdecl}{\relax}
  \BIBdecl

\bibitem{Behbahani2008}
  A.~S. Behbahani, R.~Merched, and A.~M. Eltawil, ``Optimizations of a {MIMO}
  relay network,'' \emph{IEEE Trans. Signal Process.}, vol.~56, no.~10, pp.
  5062--5073, 2008.

\bibitem{gershman2010SPM}
  A.~B. Gershman, N.~D. Sidiropoulos, S.~Shahbazpanahi, M.~Bengtsson, and
  B.~Ottersten, ``Convex optimization-based beamforming: From receive to
  transmit and network designs,'' \emph{IEEE Signal Process Mag.}, vol.~27,
  no.~3, pp. 62--75, 2010.

\bibitem{khajehnouri2007}
  N.~Khajehnouri and A.~Sayed, ``Distributed {MMSE} relay strategies for wireless
  sensor networks,'' \emph{IEEE Trans. Signal Process.}, vol.~55, no.~7, p.
  3336, 2007.

\bibitem{Rong2009TSP}
  Y.~Rong, X.~Tang, and Y.~Hua, ``A unified framework for optimizing linear
  nonregenerative multicarrier {MIMO} relay communication systems,'' \emph{IEEE
    Trans. Signal Process.}, vol.~57, no.~12, pp. 4837--4851, 2009.

\bibitem{Zeng2010Twireless}
  W.~Zeng, C.~Xiao, Y.~Wang, and J.~Lu, ``Opportunistic cooperation for
  multi-antenna multi-relay networks,'' \emph{IEEE Trans. Wireless Commun.},
  vol.~9, no.~10, pp. 3189--3199, 2010.

\bibitem{lozano2006}
  A.~Lozano, A.~Tulino, and S.~Verdu, ``Optimum power allocation for parallel
  Gaussian channels with arbitrary input distributions,'' \emph{IEEE Trans.
    Inform. Theory}, vol.~52, no.~7, pp. 3033--3051, 2006.

\bibitem{Xiao2008Gcom}
  C.~Xiao and Y.~R. Zheng, ``On the mutual information and power allocation for
  vector Gaussian channels with finite discrete inputs,'' in \emph{Proc. IEEE
    Globecom}, New Orleans, LA, Nov. 2008, pp. 1--5.

\bibitem{Zeng2010Gcom}
  W.~Zeng, M.~Wang, C.~Xiao, and J.~Lu, ``On the power allocation for relay
  networks with finite-alphabet constraints,'' in \emph{Proc. IEEE Globecom},
  Miami, FL, 2010.

\bibitem{Zeng2011TWC}
  W.~Zeng, C.~Xiao, M.~Wang, and J.~Lu, ``Linear precoding for relay networks
  with finite-alphabet inputs: Theory and practice,'' \emph{submitted to IEEE
    Trans. Wireless Commun.}, Dec. 2010.

\bibitem{palomar2006a}
  D.~Palomar and S.~Verd{\'u}, ``Gradient of mutual information in linear vector
  Gaussian channels,'' \emph{IEEE Trans. Inform. Theory}, vol.~52, no.~1, pp.
  141--154, 2006.

\bibitem{Payaro2009TIT}
  M.~Payaro and D.~P. Palomar, ``Hessian and concavity of mutual information,
  differential entropy, and entropy power in linear vector Gaussian channels,''
  \emph{IEEE Trans. Inform. Theory}, vol.~55, no.~8, pp. 3613--3628, 2009.

\bibitem{boyd2004}
  S.~Boyd and L.~Vandenberghe, \emph{Convex Optimization}.\hskip 1em plus 0.5em
  minus 0.4em\relax Cambridge University Press, 2004.

\bibitem{Manton2002TSP}
  J.~H. Manton, ``Optimization algorithms exploiting unitary constraints,''
  \emph{IEEE Trans. Signal Process.}, vol.~50, no.~3, pp. 635--650, 2002.

\bibitem{ding2007TSP}
  Y.~Ding, J.~Zhang, and K.~Wong, ``The amplify-and-forward half-duplex
  cooperative system: Pairwise error probability and precoder design,''
  \emph{IEEE Trans. Signal Process.}, vol.~55, no.~2, pp. 605--617, 2007.

\bibitem{ding2008TWC}
  ------, ``Optimal precoder for amplify-and-forward half-duplex relay system,''
  \emph{IEEE Trans. Wireless Commun.}, vol.~7, no.~8, p. 2891, 2008.

\end{thebibliography}
\end{document}